\documentclass[fleqn,10pt]{wlscirep}
\usepackage[utf8]{inputenc}
\usepackage[T1]{fontenc}
\usepackage{amsmath,amssymb}
\usepackage{siunitx}
\title{Security evaluation of quantum key distribution with weak basis-choice flaws}

\author[1,*]{Shi-Hai Sun}
\author[1]{Zhi-Yu Tian}
\author[2]{Mei-Sheng Zhao}
\author[2]{Yan Ma}
\affil[1]{School of Physics and Astronomy, Sun Yat-Sen University, Zhuhai, Guangdong 519082, P.R.China}
\affil[2]{QuantumCTek Co. Ltd., Hefei, Anhui 230000, P.R.China}

\affil[*]{sunshh8@mail.sysu.edu.cn}

\begin{abstract}
Quantum key distribution (QKD) can share an unconditional secure key between two remote parties, but the deviation between theory and practice will break the security of the generated key. In this paper, we evaluate the security of QKD with weak basis-choice flaws, in which the random bits used by Alice and Bob are weakly controlled by Eve. Based on the definition of Li \textit{et al.} [Sci. Rep. 5, 16200 (2015)] and GLLP's analysis, we obtain a tight and analytical bound to estimate the phase error and key rate for both the single photon source and the weak coherent source. Our approach largely increases the key rate from that of the original approach. Finally, we investigate and confirm the security of BB84-QKD with a practical commercial devices. 
\end{abstract}
\begin{document}

\flushbottom
\maketitle
%
%
\thispagestyle{empty}


\section*{Introduction}

Based on the principle of quantum mechanics, ``\textit{quantum cryptography}" is a possible means of implementing unconditional secure communication. One famous quantum cryptography approach is quantum key distribution (QKD) combined with One-Time pad. Since the proposal of the first QKD protocol BB84~\cite{bennett1984}, QKD has attracted much interest. The unconditional security of QKD had been proven in both perfect ~\cite{lo1999} and imperfect ~\cite{gottesman2004,lo2007} devices. QKD has also been experimentally demonstrated in fibers~\cite{tang2014a,yin2016}, free space~\cite{schmitt-manderbach2007,liao2017a}, and satellites~\cite{liao2017,liao2018}. Multi-user quantum networks based on these results have become available in many countries~\cite{elliott2005,peev2009,sasaki2011,zhang2018}.

However, because practical devices are imperfect, some assumptions of the theoretical analysis may be violated in practical situations. If the gap between theory and practice is exploited by an eavesdropper (Eve), the security of the final key may be broken. In fact, many loopholes have been discovered in practical QKD systems~\cite{lydersen2010a,li2011a,ma2013b,tang2013,sun2015,fung2007,xu2010}. These loopholes are closed by two main approaches: device-independent QKD protocols and security patches. The former include full-device-independent QKD~\cite{acin2007,pironio2009}, measurement-device-independent QKD~\cite{lo2012,ma2012,wang2016}, and semi-device-independent QKD~\cite{pawlowski2011}. Security patches account for the parameters of practical devices (as many as possible) in the security model. Although device-independent QKD can remove \textit{all or a portion} of the loopholes, the task remains technologically challenging, especially in practical commercial QKD networks. Thus, most practical QKD systems implement security patches. 

In the BB84 protocol, both Alice and Bob must determine how to prepare and measure the quantum states. For this purpose, they require random bits. In practical situations, the random bits may be weakly known or controlled by Eve, and the security of the generated key is compromised. A typical attack that exploits the weak randomness of QKD is wavelength attack~\cite{li2011a,ma2013b}. The security of QKD with weak randomness was first studied by Li \textit{et al.}~\cite{li2015}, and has since been applied to different cases~\cite{li2018,zhang2019}. In Li's analysis, if the legitimate parties use the ``one-step post processing method" to distill the final key, even a small degree of non-randomness will rapidly reduce the final key rate. The key rate can be improved if the legitimate parties adopt the ``two-step post processing method", or biased basis protocol, in which Alice and Bob distill the key from the rectilinear and diagonal bases, respectively. However, to maximize Eve's information, they must perform global optimization, which is hampered by at least two disadvantages: large time cost and convergence to a local optimum. The time cost is incurred by the complexity or cost of post processing, and local (rather than global) optimization compromises the security of the generated key.

To mitigate these problems, we develop an analytical formula that estimates the key rate for both the single photon source (SPS) and the weak coherent source (WPS). In numerical simulations, our method significantly increased the key rate over the original method of Li \text{et al.}~\cite{li2015}. For example, the original method of Ref.~\cite{li2015} can generate no secure key for a SPS with a basis-choice flaw of 0.1 when the bit error rate exceeds 3.4\%, but our method achieves a final key rate of 0.45 under these conditions. Furthermore, to evaluate the performance of QKD under wavelength attack, we also estimate the key rate of a practical QKD system with a passive basis-choice.

\section*{Results}

\subsection*{Weak randomness and one-step post processing}\label{sec:weak_randoness}
This section briefly reviews the analysis of Ref.~\cite{li2015}. Alice determines her quantum state from two random bits: $x_0$ for bit and $x_1$ for basis. Meanwhile, Bob chooses his basis from a random bit, $y$. As the final key is distilled only when Alice and Bob choose the same basis ($x_1=y$), the following analysis is limited to the case $x_1=y$. In a practical QKD system, $x_0$ ($x_1$) may be weakly controlled by Eve with a hidden variable $\lambda_0$ ($\lambda_1$). Setting $k$ and $k'$ = [0, 1] as the values of $x_0$ and $x_1$, respectively, the probabilities of obtaining $x_0=k$ and $x_1=k'$ are respectively given by
\begin{equation}
	\begin{split}
		p\left(x_{0}=k\right)&=\sum_{i} p_{\lambda_{0}=i} p\left(x_{0}=k | \lambda_{0}=i\right), \\p\left(x_{1}=k'\right)&=\sum_{j} p_{\lambda_{1}=j} p\left(x_{1}=k' | \lambda_{1}=j\right).
	\end{split}
\end{equation}
Here $\sum_{i} p_{\lambda_{0}=i}=\sum_{j} p_{\lambda_{1}=j}=1$. Due to the existence of the hidden variable $\lambda_0$, we cannot guarantee that $p(x_0=0|\lambda_{0}=i)=p(x_0=1|\lambda_{0}=i)=1/2$ holds for all $i$, even if $p(x_0=0) = p(x_0 =1)=1/2$ holds. The same conclusion is reached for $p(x_1=k')$. To evaluate the weak randomness of $x_0$ and $x_1$, the deviations is defined as
\begin{equation}
	\begin{split}
		\left|p\left(x_{0}=k | \lambda_{0}=i\right)-\frac{1}{2}\right| 
		\leq \varepsilon_{0}, \\ \left| p\left(x_{1} =k' | \lambda_{1}=j\right)-\frac{1}{2} 
		\right|\leq \varepsilon_{1},
	\end{split}
\end{equation}
respectively. 
Here, $0 \leq \varepsilon_0, \varepsilon_1 \leq 1$ define the amount of prior information known to Eve. 

When the hidden variables $\lambda_{0}=i$ and $\lambda_{1}=j$ are given, the quantum state shared by Alice and Bob can be written as~\cite{li2015}
\begin{equation}\label{rho_AB_ij}
	\begin{split} 
		\rho_{A B}^{i,j}=& \sum_{\mu, \nu} q_{\mu, \nu}\left\{p\left(x_{1}=0 | \lambda_{1}=j\right) I \otimes X^{\mu} Z^{\nu}|\varphi\rangle\left\langle\left.\varphi\right|_{\lambda_{0}=i} Z^{\nu} X^{\mu} \otimes I\right.\right.\\ &+p\left(x_{1}=1 | \lambda_{1}=j\right) I \otimes H X^{\mu} Z^{\nu} H|\varphi\rangle\left\langle\left.\varphi\right|_{\lambda_{0}=i} H Z^{\nu} X^{\mu} H \otimes I\right\}, 
	\end{split}
\end{equation}
with
\begin{equation}
	|\varphi\rangle_{\lambda_{0}=i}=\sqrt{p\left(x_{0}=0 | 
		\lambda_{0}=i\right)}|00\rangle+\sqrt{p\left(x_{0}=1 | 
		\lambda_{0}=i\right)}|11\rangle.
\end{equation}
Here, $\mu,\nu \in \{0,1\}$, and $q_{\mu,\nu}$ is the probability that Eve performs different operators on the quantum state $|\varphi\rangle_{\lambda_{0}=i}$. These probabilities satisfy $\sum_{\mu,\nu} q_{\mu,\nu}=1$. $I$ is the unity matrix, $X$ and $Z$ are Pauli matrices, and $H=\frac{1}{\sqrt{2}}\begin{bmatrix}1 &1\\1 & -1\end{bmatrix}$.

In the following, we first discuss the case of SPS, then expand our results to WPS. In SPS, the total key rate is given by
\begin{equation}\label{keyrate}
	R \geq \sum_{i,j} p_{\lambda_0=i} p_{\lambda_{1}=j} R^{i,j}.
\end{equation}
The key rate for a given $i,j$ is
\begin{equation}\label{keyrate_ij_sps}
	R^{i,j}_{sps} \geq 1-H(e_{phase}^{i,j})- H(e_{bit}^{i,j}),
\end{equation}
where $e_{bit}^{i,j}$ ($e_{phase}^{i,j}$) is the bit error (phase error) of the given $i,j$. Because an experiment reveals only the total bit error $e_{bit} 
= \sum_{i,j}p_{\lambda_0 =i}p_{\lambda_1 =j} e_{bit}^{i,j}$, Eq.~\ref{keyrate} can be rewritten as
\begin{equation}\label{keyrate_sps}
	\begin{split}
		R_{sps} &= \sum_{i,j} p_{\lambda_0=i} p_{\lambda_{1}=j} \left[ 1-H(e_{phase}^{i,j})- H(e_{bit}^{i,j})\right] \\
		&\geq 1 -H(\sum_{i,j}p_{\lambda_0=i} p_{\lambda_{1}=j} e_{phase}^{i,j}) -H(e_{bit})\\
		&\equiv 1-H(e_{phase})- H(e_{bit}).
	\end{split}
\end{equation} 
Here $e_{phase}$ is the total phase error, and the second inequality uses the Jensen inequality because $H(x)$ is concave. Before obtaining the lower bound of the key rate, we should estimate the upper bound of $e_{phase}$. The authors of ~\cite{li2015} proved that for the density matrix given by Eq.~\ref{rho_AB_ij}, the upper bound of the phase error can be written as
\begin{equation}\label{phase_error_originalmethod}
	\begin{split}
		e_{phase}^{i,j}& \leq e_{bit}^{i,j} +\delta, \\
		e_{phase}= \sum_{i,j}p_{\lambda_0=i} p_{\lambda_{1}=j} e_{phase}^{i,j} &\leq \sum_{i,j}p_{\lambda_0=i} p_{\lambda_{1}=j} e_{bit}^{i,j} +\delta =e_{bit} +\delta,
	\end{split}
\end{equation}
where
\begin{equation}\label{delta_01_define}
	\delta=\max \left\{\delta_0 \equiv \frac{1}{2}- 
	\sqrt{\frac{1}{4}-\varepsilon_0^2},\quad  
	\delta_1 \equiv 2\varepsilon_1\right\}.
\end{equation}

The final key rate of a SPS (Eq.~\ref{keyrate_sps}) is then rewritten as
\begin{equation}\label{keyrate_sps_o}
	R^o_{sps} \geq 1-H(e_{bit} +\delta) -H(e_{bit}).
\end{equation}
In this expression, the superscript $o$ distinguishes the original method from our proposed method, which is introduced later. Most practical QKD systems use a WPS. Following GLLP's analysis~\cite{gobby2004}, the key rate of Eq.~\ref{keyrate} is then written as
\begin{equation}\label{keyrate_wps_o}
	\begin{split}
		R^{o}_{wps} & \geq \sum_{i,j}p_{\lambda_{0}=i}p_{\lambda_{1}=j} R^{i,j} \\ & \geq \sum_{i,j}p_{\lambda_{0}=i}p_{\lambda_{1}=j} \left\{-Q_s^{i,j} f(E_{s,bit}^{i,j})H(E_{s,bit}^{i,j}) +Q_{s,1}^{i,j} \left[1-H(e_{1,bit}^{i,j}+\delta)\right]\right\}\\
		&\geq -Q_s f(E_s) H(\sum_{i,j}p_{\lambda_{0}=i}p_{\lambda_{1}=j} E_{s,bit}^{i,j}) +Q_{s,j} \left[1 -H(\sum_{i,j}p_{\lambda_{0}=i}p_{\lambda_{1}=j} e_{1,bit}^{i,j}+\delta)\right]\\
		&\equiv
		-Q_s f(E_{s,bit}) H(E_{s,bit}) +Q_{s,1} [1- H(e_{1,bit}+\delta)].
	\end{split}
\end{equation}
Here the subscript $s$ denotes the key generated from the signal state with intensity $s$. $Q_s^{i,j}$ ($E_{s,bit}^{i,j}$) is the total gain (bit error) for a given $i,j$, and $Q_{s,1}^{i,j}$ ($e_{1,bit}^{i,j}$) is the yield (bit error) of the single photon pulse for the given $i,j$. The terms $Q_s=\sum_{i,j} p_{\lambda_{0}=i}p_{\lambda_{1}=j} Q_{s}^{i,j}$ ($Q_{s,1}=\sum_{i,j} p_{\lambda_{0}=i}p_{\lambda_{1}=j} Q_{s,1}^{i,j}$) and $E_{s,bit}=\sum_{i,j} p_{\lambda_{0}=i}p_{\lambda_{1}=j} E_{s,bit}^{i,j}$ ($e_{1,bit}=\sum_{i,j} p_{\lambda_{0}=i}p_{\lambda_{1}=j} e_{1,bit}^{i,j}$) are the total gain and error, respectively, for all $i,j$.  $f(E_{s,bit})=f(E_{s,bit}^{i,j})=1.22$ is the efficiency of the error correction, which can be considered constant. In the third inequality, we recognize that $Q_s \geq Q_{s}^{i,j} \geq 0$ and $Q_{s,1} \geq Q_{s,1}^{i,j} \geq 0$ for all $i,j$. The gain $Q_s$ and error $E_{s,bit}$ in the equality can be directly measured in experiments, and the contributions of the single photon pulse ($Q_{s,1}$ and $e_{1,bit}$) should be estimated by the decoy state method~\cite{hwang2003,lo2005,wang2005a}.

\subsection*{Our method with one-step post processing}\label{sec:our_method_onestep}
In this section, we show that the upper bound of the phase error (Eq.~\ref{phase_error_originalmethod}) is suboptimal, and that the key rate can be improved by imposing a tight bound. Given the density matrix $\rho_{AB}^{i,j}$ (Eq.~\ref{rho_AB_ij}), the bit error rate and phase error rate are respectively written as 
\begin{equation}
	\begin{split}
		e_{bit}^{i,j}&=\langle \phi_2 |\rho_{AB}^{i,j}|\phi_2 \rangle +\langle \phi_4 |\rho_{AB}^{i,j}|\phi_4 \rangle =q_{01} p(x_1=1|\lambda_1=j) +q_{10} p(x_1=0|\lambda_1=j) 
		+q_{11}\\ & \geq \left(\frac{1}{2} -\varepsilon_1 \right) (q_{01}+q_{10}) +q_{11} \equiv e_{bit,low}^{i,j}, \\
		e_{phase}^{i,j} &=\langle \phi_3 |\rho_{AB}^{i,j}|\phi_3 \rangle +\langle \phi_4 |\rho_{AB}^{i,j}|\phi_4 \rangle \\ 
		& = q_{00} \delta_0 +q_{11} (1-\delta_0)+ q_{01} 
		\left\{\frac{1}{2}+ \left[p(x_1=0|\lambda_1=j)- p(x_1=1|\lambda_1=j)  \right](\frac{1}{2}-\delta_0) \right\}\\
		& +q_{10} \left\{\frac{1}{2}  +[p(x_1=1|\lambda_1=j) -p(x_1=0|\lambda_1=j) ](\frac{1}{2}-\delta_0) \right\}.
	\end{split}
\end{equation}
Here, $|\phi_1 \rangle =(|00\rangle +|11\rangle)/\sqrt{2}$, $|\phi_2 \rangle =(|01\rangle +|10\rangle)/\sqrt{2}$,$|\phi_3 \rangle =(|00\rangle - |11\rangle)/\sqrt{2}$,$|\phi_4 \rangle =(|01\rangle - |10\rangle)/\sqrt{2}$ are the four Bell states. Thus we have
\begin{equation}
	\begin{split}
		e_{phase}^{ij}-e_{bit}^{i,j} & = q_{00} \delta_0 -q_{11}\delta_0 +(q_{01}-q_{10}) (1- \delta_0)[2p(x_1=0|\lambda_1 =j) -1]\\
		& \leq q_{00} \delta_0 +q_{11} \delta_0 + 2\varepsilon_1 q_{01} +2\varepsilon_1 q_{10} \\& = (q_{00} +q_{11})\delta_0 +2\varepsilon_1 \frac{e_{bit,low}^{i,j} -q_{11}} 
		{1/2 -\varepsilon_1}\\	&=\frac{4\varepsilon_1}{1-2\varepsilon_1} e_{bit,low}^{i,j} -q_{11} \frac{4\varepsilon_1}{1-2\varepsilon_1} +(q_{00} +q_{11})\delta_0 \\	&\leq \frac{4\varepsilon_1}{1-2\varepsilon_1} e_{bit}^{i,j} +\delta_0,
	\end{split}
\end{equation}
where $\delta_0$ is defined in Eq.~\ref{delta_01_define}, and $0 \leq q_{\mu\nu} \leq 1$ for all $q_{\mu\nu}$. Thus, the upper bound of the phase error can be written as
\begin{equation}\label{phase_error_ourmethod}
	e_{phase}^{i,j} \leq \frac{1+ 2\varepsilon_1}{1-2\varepsilon_1} 
	e_{bit}^{i,j} +\frac{1}{2} -\sqrt{\frac{1}{4} -\varepsilon_0^2}.
\end{equation}
Submitting the above inequality into Eq.~\ref{keyrate} and applying the method described in Sec.~\ref{sec:weak_randoness}, the final key rate is rewritten as
\begin{equation}\label{keyrate_ourmethod}
	R^{t} \geq
	\begin{cases}
		1 - H\left(\frac{1+2\varepsilon_1}{1-2\varepsilon_1} e_{bit} + \delta_0\right) -H(e_{bit}) & \text{for SPS}\\
		-Q_s f(E_{s,bit}) H(E_{s,bit}) +Q_{s,1} \left[1- H\left(\frac{1+2\varepsilon_1}{1-2\varepsilon_1} e_{1,bit} +\delta_0 \right) \right] & 
		\text{for WPS}
	\end{cases}.
\end{equation} 

\begin{figure}[ht]
	\centering
	\includegraphics[width=8cm]{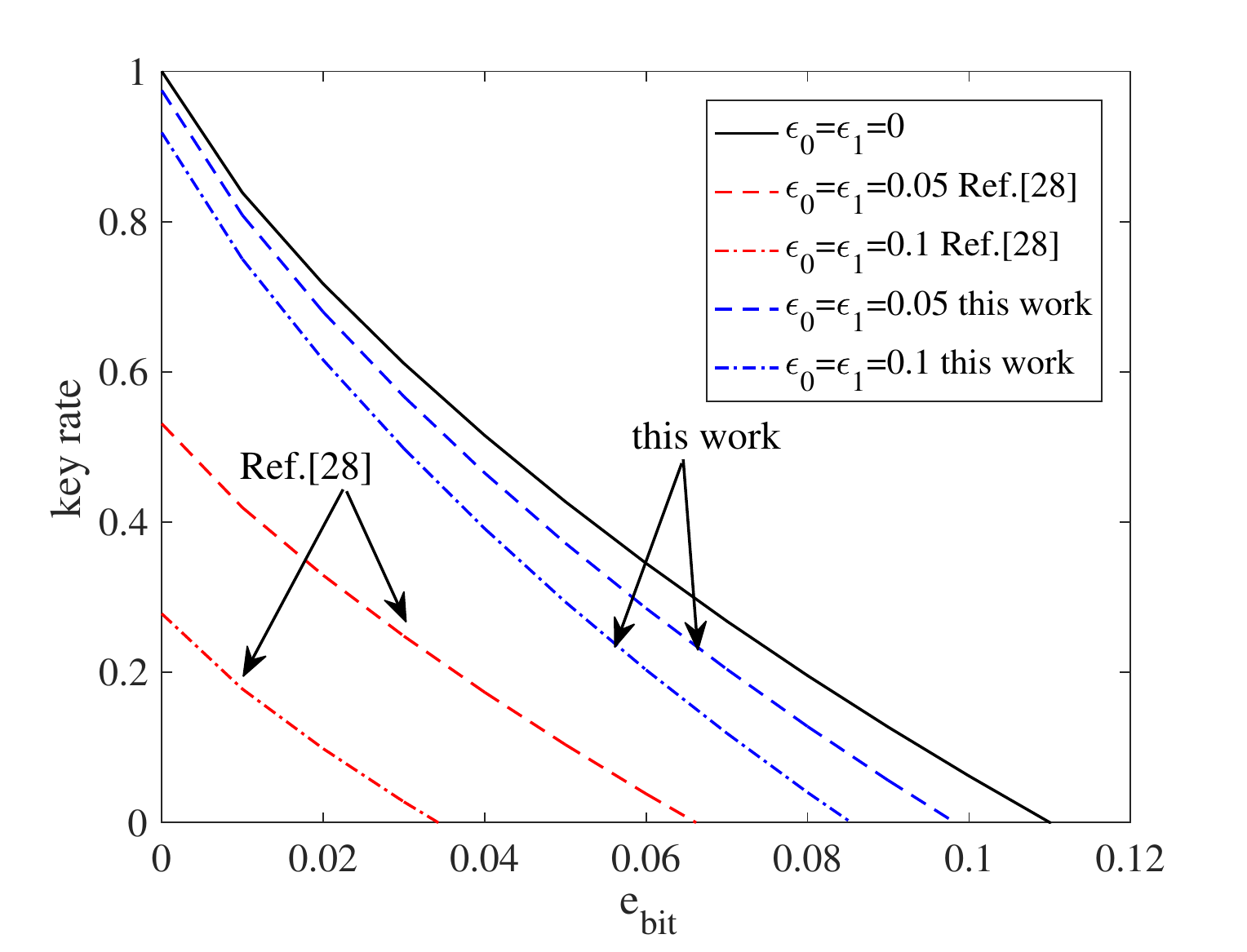}
	\includegraphics[width=8cm]{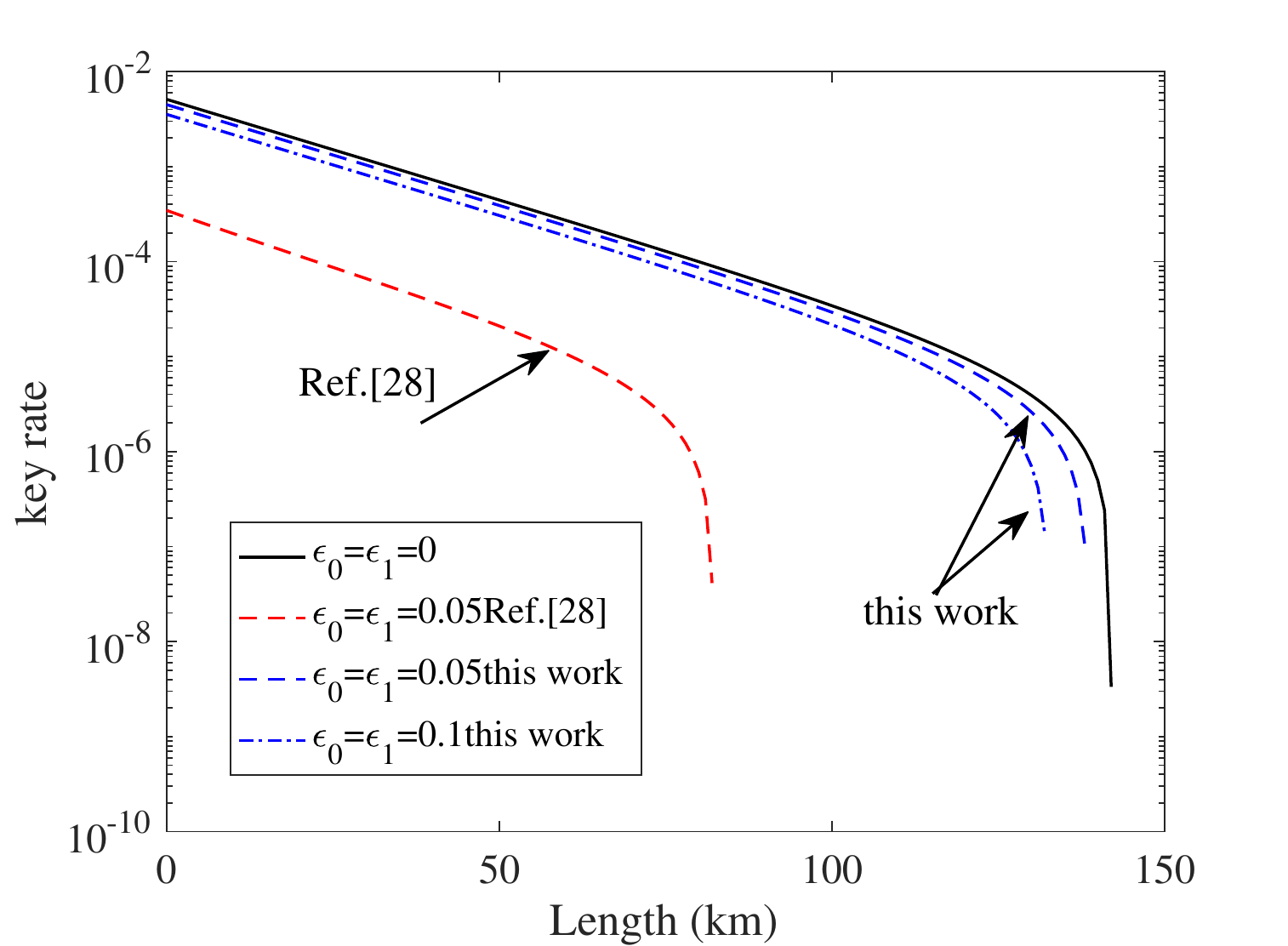}	
	\caption{Key rates in the original analysis ~\cite{li2015} (red lines) and the method proposed in this paper (blue lines). Results are plotted for SPS (left) and WPS (right). The black solid line is the result of the ideal case without basis-choice flaws. To simplify the simulation, we assume $\varepsilon_0 =\varepsilon_1$ and infinite decoy states. The WPS case employs the experimental results of GYS~\cite{gobby2004}; thus, the signal state intensity is $s=0.48$ and the other parameters are set as follows: dark count rate $Y_0 =1.7 \times 10^{-6}$, background error rate $e_0=0.5$, fiber loss 0.21 \si{dB}/\si{\kilo\meter}, Bob's transmittance $\eta_{Bob}=0.045$, and error rate of optical devices $e_{det}=3.3\%$. The method of Ref.~\cite{li2015} generates no key in the case of WPS with $\varepsilon_0 =\varepsilon_1 =0.1$.}
	\label{fig:one_step_com}
\end{figure}

Fig.~\ref{fig:one_step_com} compares the numerical simulation results of the method in Ref.~\cite{li2015} (Eq.~\ref{keyrate_sps_o} and Eq.~\ref{keyrate_wps_o}) and our method (Eq.~\ref{keyrate_ourmethod}). Our method significantly improved the key rate for both SPS and WPS (the simulation method is given in Appendix~\ref{appendix_sec:simulation}). For example, in the SPS case with $\varepsilon_0=\varepsilon_1=0.1$, the maximal tolerable error rate was only 3.4\% in the method of Ref.~\cite{li2015}, but was increased to 8.5\% by our method. In the WPS case with $\varepsilon_0=\varepsilon_1=0.1$, no secure key was generated by the method in Ref.~\cite{li2015}, but a final key of fiber length 132 \si{\kilo\meter} was generated by our method. 

To evaluate the security of practical QKD with weak randomness flaws, we test the performance of commercial BS which may suffer from the wavelength attack. The experimental scheme and results are given in Appendix~\ref{sec:experiment}, and the estimated key rate is listed in Tab.~\ref{tab:experimental_result}.

\begin{table}	
	\caption{Key rates estimated by our method. Here $\gamma =(R_{ideal}-R_{prac})/R_{ideal}$defines the practical key rate ($R_{prac}$) relative to the ideal key rate without basis-choice flaws ($R_{ideal}$). In the simulations, we set $\varepsilon_0 =0$, and the other parameters were those assumed in Fig.~\ref{fig:one_step_com}.}\label{tab:experimental_result}
	\renewcommand*{\arraystretch}{1.15}
	\begin{tabular}[t]{c|ccccc}
		\hline\hline
		T & $-5^\circ$ \qquad &$18^\circ$ \qquad &$25^\circ$ \qquad & 
		$40^\circ$ \qquad 
		&$70^\circ$ \\
		\hline
		$\varepsilon_1$ & \qquad 0.0253 &\qquad 0.0235 &\qquad 0.0250 &\qquad 
		0.0274 & \qquad 0.0275\\
		$\gamma$ (SPS)$@ e_{bit}=3\%$ & 0.0259 & 0.0239 & 0.0255 & 0.0281 & 
		0.0282 \\
		$\gamma$ (SPS)$@ e_{bit}=5\%$ & 0.0521 & 0.0482 & 0.0514 & 0.0566 & 
		0.0567 \\
		$\gamma$ (SPS)$@ e_{bit}=7\%$ & 0.1017 & 0.0941 & 0.1004 & 0.1105 & 
		0.1108\\
		$\gamma$ (WPS)$@ L=10 \si{\kilo\meter}$& 0.0445 & 0.0411 & 0.0438 & 
		0.0483 & 0.0484 \\
		$\gamma$ (WPS)$@ L=50 \si{\kilo\meter}$& 0.0454 & 0.0420 & 0.0448 & 
		0.0493 & 0.0495 \\
		$\gamma$ (WPS)$@ L=100 \si{\kilo\meter}$& 0.0548 & 0.0507 & 0.0541 & 
		0.0596 & 0.0597 \\
		\hline\hline
	\end{tabular}
\end{table}

\subsection*{Biased base QKD protocol}\label{sec:biased-bases}
In some practical QKD systems, two bases ($Z$ and $X$) can deliver different gain or error rate performances. Therefore, to improve the total key rate, we let Alice and Bob observe bases $Z$ and $X$, respectively. In this section, we analyze the security of biased base QKD with weak randomness. When Alice and Bob distill the key from their respective bases, the key rate becomes
\begin{equation}\label{keyrate_biased_bases_sps}
	R^{two}_{SPS} \geq p_{rec}[1-H(e_{rec}^b) -H(e_{rec}^p)] +p_{dia} [1- 
	H(e_{dia}^b) -H(e_{dia}^p)].
\end{equation}
Here $e_{rec}^b$ and $e_{rec}^p$ ($e_{dia}^b$ and $e_{dia}^p$) are the bit error and phase error rates, respectively, in the rectilinear (diagonal) basis. Their values are given by
\begin{equation}\label{bit_phase_error_rate}
	\begin{split}
		e_{rec}^p &= \frac{p_{rec1} e_{b 01}+p_{rec2} e_{b 11}}{p_{rec}} 
		+\frac{1}{2}-\sqrt{-\epsilon_{0}^{2}+\frac{1}{4}}, \\
		e_{dia}^p &= \frac{p_{\text {dial}} e_{b 00}+p_{\text {dia} 2} e_{b 10}}{p_{\text{dia}}}+\frac{1}{2}-\sqrt{-\epsilon_{0}^{2}+\frac{1}{4}}\\e_{rec}^b &=\frac{p_{\text {rec1}} e_{b 00}+p_{\text {rec} 2} e_{b 10}}{p_{\text {rec}}}, \\e_{dia}^b &=\frac{p_{\text {dial}}, e_{b 01}+p_{\text {dia} 2} e_{b 11}}{P_{\text {dia}}}.
	\end{split}
\end{equation}
In these expressions, $e_{b00}$ and $e_{b01}$ ($e_{b10}$ and $e_{b11}$) are the bit error rates in the rectilinear and diagonal bases, respectively, given a hidden variable $\lambda_1=0$ ($\lambda_1 =1$). $p_{rec}$ and $p_{dia}$ are the probabilities that Bob obtains the outcome in the rectilinear and diagonal bases, respectively. They are calculated as
\begin{equation}
	p_{rec}=p_{rec1}+ p_{rec2}, \quad p_{dia} =p_{dia1}+ p_{dia2},
\end{equation} 
respectively, where $p_{rec1}=p_{\lambda_{1}=0} p\left (x_{1}=0 | \lambda_{1}=0\right)$, 
$p_{rec2}=p_{\lambda_{1}=1} p\left(x_{1}=0 | \lambda_{1}=1\right)$, 
$p_{dia1}=p_{\lambda_{1}=0} p\left(x_{1}=1 | 
\lambda_{1}=0\right)$, and $p_{dia2}=p_{\lambda_1=1} p\left (x_1=1|\lambda_1=1 
\right)$. 

As the bit error rates $e_{rec}^b$ and $e_{dai}^b$ can be directly measured in experiments, we need only to estimate the upper bounds of the phase error rates $e_{rec}^p$ and $e_{dai}^p$. Using Eq.~\ref{bit_phase_error_rate}, the phase error $e_{rec}^p$ becomes
\begin{equation}
	\begin{split}
		e_{rec}^p &=e_{dia}^b \frac{p_{dia}}{p_{rec}}\frac{p_{rec1}e_{b01} 
			+p_{rec2}e_{b11}}{p_{dia1}e_{b01} +p_{dia2}e_{b11}} +\delta_0 \\
		&= e_{dia}^b \frac{p_{dia}}{p_{rec}}\frac{p_{\lambda_1=0} p(x_1=0|\lambda_1=0) 
			e_{b01} +p_{\lambda_1=1} p(x_1=0|\lambda_1=1) e_{b11}}{p_{\lambda_1=0} 
			p(x_1=1|\lambda_1=0) e_{b01} +p_{\lambda_1=1} p(x_1=1|\lambda_1=1)e_{b11}} 
		+\delta_0 \\
		&\leq e_{dia}^b \frac{p_{dia}(1/2 +\varepsilon_1)}{p_{rec}(1/2 
			-\varepsilon_1)}  +\delta_0. 
	\end{split}
\end{equation}
By the above method, we also obtain
\begin{equation}
	e_{dia}^p \leq  e_{rec}^b \frac{p_{rec}(1/2 
		+\varepsilon_1)}{p_{dia}(1/2 -\varepsilon_1)}  +\delta_0.
\end{equation}
Here we assume that the bit weak randomness in Z-basis is the same as that of X-basis, thus the same $\delta_0$ is used in the equations above. But, by considering $\delta_0$ in the bases respectively, our analysis is also valid for the QKD system with different bit weak randomness.Then Eq.~\ref{keyrate_biased_bases_sps} can then be written as
\begin{equation}\label{keyrate_twoway_sps}
	\begin{split}
		R^{two}_{SPS} &\geq p_{rec} \left\{1-H(e_{rec}^b) -H\left[e_{dia}^b 
		\frac{p_{dia}(1/2 +\varepsilon_1)}{p_{rec}(1/2 -\varepsilon_1)}  
		+\delta_0\right] \right\} \\ & +p_{dia} \left\{1-H(e_{dia}^b) -H\left[e_{rec}^b 
		\frac{p_{rec}(1/2 +\varepsilon_1)}{p_{dia}(1/2 -\varepsilon_1)}  
		+\delta_0\right] \right\}.	
	\end{split}
\end{equation}

Applying the method above and GLLP's analysis, the key rate given by Eq.~\ref{keyrate_twoway_sps} can be expanded to the WPS case as follows: 
\begin{equation}\label{keyrate_twoway_wps}
	\begin{split}
		R_{wps}^{two}& \geq p_{rec}\left\{ - Q_{s,rec} f(E_{s,rec}^b) H(E_{s,rec}^b) +Q_{1,rec}\left[1 - H(e_{1,dia}^b \frac{p_{dia}(1/2+\varepsilon_{0})}{p_{rec}(1/2-\varepsilon_{1})} +\delta_0)\right]\right\}\\
		&+ p_{dia}\left\{ - Q_{s,dia} f(E_{s,dia}^b) H(E_{s,dia}^b) +Q_{1,dia}\left[1 - H(e_{1,rec}^b \frac{p_{rec}(1/2+\varepsilon_{0})}{p_{dia}(1/2-\varepsilon_{1})} +\delta_0)\right]\right\},
	\end{split}
\end{equation}
where $Q_{s,rec}$ ($Q_{s,rec}$) and $E_{s,dia}^b$ ($E_{s,dia}^b$) denote the total gain and bit error rate, respectively, in the rectilinear (diagonal) base, and $Q_{1,rec}$ ($Q_{1,rec}$) and $e_{1,dia}^b$ ($e_{1,dia}^b$) are the gain and bit error rate, respectively, of a single photon pulse in the rectilinear (diagonal) base.

\section*{Discussion}
We evaluated the security of QKD with weak basis-choice flaws. The previous analysis of Li \textit{et al.}~\cite{li2015} was extended by applying a tight analytical bound for estimating the phase error. The final key rate was significantly improved by the proposed approach. For example, when $\varepsilon_0=\varepsilon_1=0.1$ and the bit error rate exceeded 3.4\%, no final key was generated by the previous method, but a final key rate of 0.45 was achieved by our method. Applying our analysis, we evaluated the security of a practical QKD system in which Bob passively chooses his basis with a BS. In experiments using a practical BS with typical parameters, the key rate was reduced by less than 6\%. Thus, the proposed method improves the QKD performance even in weak randomness scenarios.

Note that, although we analyze the weak randomness of basis-choice in this paper, there are other imperfections in source and detection. Thus, how to take all of these imperfections in one general mode is still an open question, and we will discuss it in our further works

\section*{Methods}

\subsection*{Formulations of simulation}\label{appendix_sec:simulation}
This Appendix shows the simulation formulations of Fig.~\ref{fig:one_step_com} and Table~\ref{tab:experimental_result}. In the absence of Eve, the total gain and error rate are respectively written as
\begin{equation}
\begin{split}
Q_\omega &= 1-(1-Y_0)e^{-\omega \eta}\\
and
E_\omega Q_\omega& =e_0 Y_0 +e_{det}(1-e^{-\omega \eta}).
\end{split}
\end{equation}
Here, $\omega \in \{s\}$ and $\omega \in \{d\}$ are the intensities of the signal state ($s$) and decoy state ($d$), respectively, $Y_0$ is the dark count of the single photon detector, and $e_0$ is the background error rate. $\eta$ is the total transmittance of the system, which is given by
\begin{equation}
\eta =\eta_{Bob} 10^{-\alpha /10}.
\end{equation}
In this expression, $\eta_{Bob}$ is the transmittance of Bob's optical devices and the efficiency of single- photon detectors, and $\alpha$ is the channel loss. In QKD with a weak coherent source, photon-number-dependent attacks (such as photon-number splitting attacks) must be removed by the decoy state method. Assuming that Alice and Bob use infinite decoy states, the gain and quantum bit error rate of a single photon pulse are respectively given by
\begin{equation}
\begin{split}
Q_{s,1} & = s e^{-s} [1- (1-Y_0)(1-\eta)] \equiv s e^{-s} Y_1,\\
e_1 & =(e_0 Y_0 +e_{det} \eta)/ Y_1.
\end{split}
\end{equation}

\subsection*{Experiment}\label{sec:experiment}
\begin{figure}[ht]
	\centering
	\includegraphics[width=6cm]{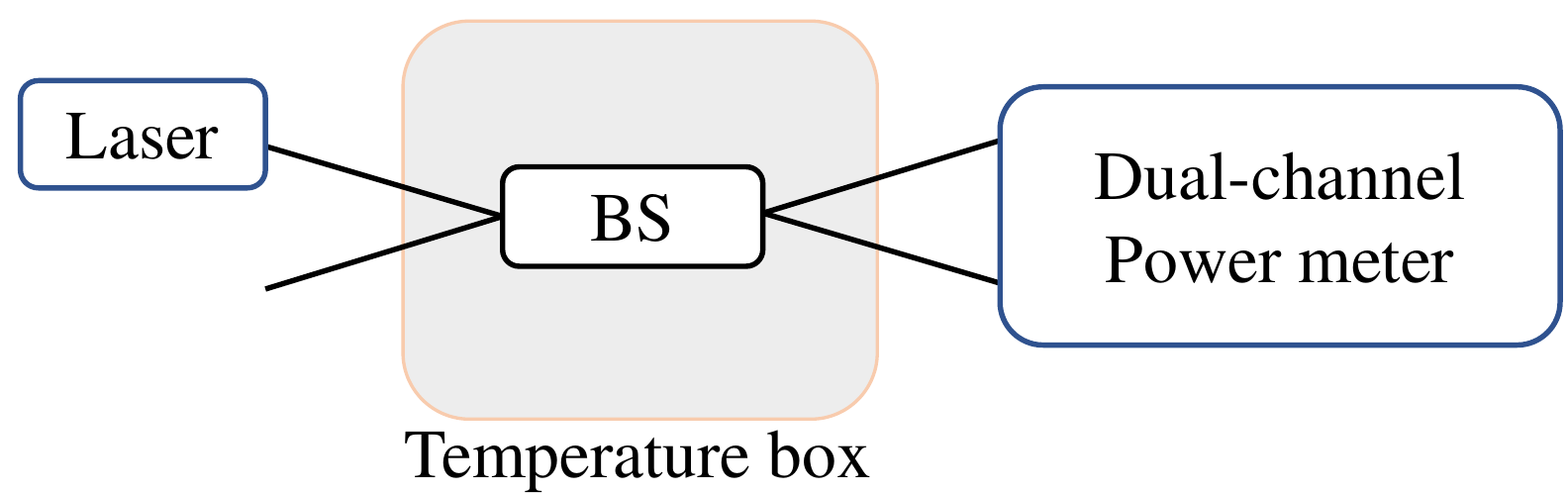}	
	\includegraphics[width=8cm]{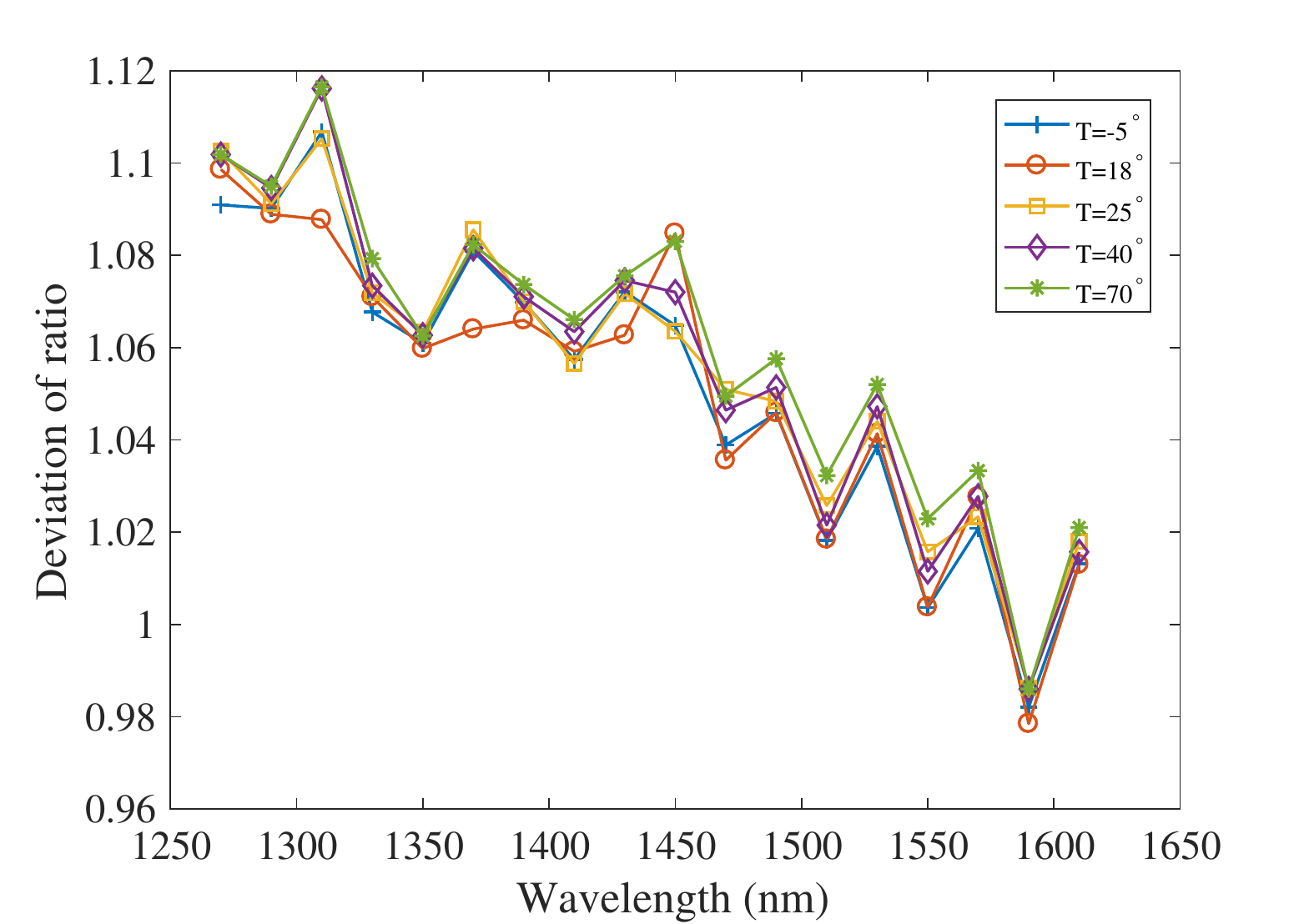}
	\caption{Experimental scheme (left) and measured deviation ratio (right) of a practical commercial BS that evaluates the key rate under a weak measured basis flaw. The deviation ratio is determined by Eq.~\ref{deviation_ratio} and $T$ is the working temperature of the BS, which is controlled by the temperature box.}
	\label{fig:experimental}
\end{figure} 

In some practical QKD systems, Bob passively chooses his measured basis with a BS. Because this scheme requires no active modulator, it enables high-speed, low-cost, and low-complexity operations. However, (as is well known) the transmittance of the BS may depend on the wavelength of the light, opening a potential loophole for wavelength attack by Eve~\cite{li2011a,ma2013b}. In this section, we evaluate the performance of QKD with a passive BS by the above analysis.

The experimental scheme is shown in the left panel of Fig.~\ref{fig:experimental}. The BS was encased in a temperature box that controlled its working temperature. The input of the BS was a tunable laser (model JW3113; province, country), and the light output was measured by a dual-channel power meter (model JW8103D, province, country). In a perfect BS, the measured power of both power meters is identical. The performance of a real BS is defined by its deviation ratio as follows:
\begin{equation}\label{deviation_ratio}
\Delta(T,\lambda) = \frac{P_0(T,\lambda)}{P_1(T,\lambda)}.
\end{equation}
Here, $P_0$ and $P_1$ are the powers of the light measured by the two optical power meters, $T$ is the working temperature of the BS, and $\lambda$ is the wavelength of the input light. 

The measured deviation ratio of the BS was measured at different wavelengths of the input light and different working temperatures. The results are shown in the right panel of Fig.~\ref{fig:experimental}. From the experimental results, the weak randomness in Bob's basis-choice can be estimated as
\begin{equation}
\varepsilon_1 =\max_{\lambda} \left|\frac{P_0}{P_0 +P_1} -\frac{1}{2} 
\right|= \max_{\lambda} \left| \frac{\Delta}{1+ 
	\Delta} -\frac{1}{2} \right|.
\end{equation} 
The estimated key rates in the SPS and WPS cases are listed in Table~\ref{tab:experimental_result} (see main text). At communication distances smaller than 100 \si{\kilo\meter}, the key rate was reduced by less than 6\%. Topical subheadings are allowed. Authors must ensure that their Methods section includes adequate experimental and characterization data necessary for others in the field to reproduce their work.

\bibliography{library}

\section*{Acknowledgements (not compulsory)}

The authors thank H.W. Li for helpful discussions on the simulation. This work was supported by the National Natural Science Foundation of China (NSFC) (11674397).

\section*{Author contributions statement}

Shi-Hai Sun and Mei-Sheng Zhao finish the theoretical analysis. Zhi-Yu Tian gets the simulation with the help of Shi-Hai Sun. Yan Ma conducted the experiment. All authors reviewed the manuscript. 

\section*{Additional information}

There is no interesting conflict.

\end{document}